\documentclass{elsart}

\begin{document}
\begin{frontmatter}
\title{A succinct presentation of the quantized Klein-Gordon field,
       and a similar quantum presentation of the classical Klein-Gordon random field}
\author{Peter Morgan}
\address{Physics Department, Yale University.}
\ead{peter.w.morgan@yale.edu}
\ead[url]{http://pantheon.yale.edu/$\sim$PWM22}

\begin{abstract}
A succinct presentation of the algebraic structure of the quantized Klein-Gordon field
can be given in terms of a Lorentz invariant inner product.
A presentation of a classical Klein-Gordon \emph{random} field at non-zero temperature
can be given in the same noncommutative algebraic style, allowing a detailed comparison
of the quantized Klein-Gordon field with a classical Klein-Gordon random field.
\end{abstract}

\begin{keyword}
quantum field theory \sep measurement \sep classical random fields
\PACS 03.70.+k \sep 03.65.Bz \sep 05.40.-a
\end{keyword}
\end{frontmatter}

\newcommand\Half{{\frac{1}{2}}}
\newcommand\Intd{{\mathrm{d}}}
\newcommand\eqN{{\,\stackrel{\mathrm{N}}{=}\,}}
\newcommand\PP[1]{{(\hspace{-.27em}(#1)\hspace{-.27em})}}
\newcommand\PPs[1]{{(\hspace{-.4em}(#1)\hspace{-.4em})}}
\newcommand\RR {{\mathrm{I\hspace{-.1em}R}}}
\newcommand\CC{{{\rm C}\kern -0.5em 
          \vrule width 0.05em height 0.65em depth -0.03em
          \kern 0.45em}}
\newcommand\kT{{{\mathsf{k_B}} T}}
\newcommand{\RA}{\mathcal{R}_A}
\newcommand{\RB}{\mathcal{R}_B}
\newcommand{\PastRA}{\textsl{Past}(\RA)}
\newcommand{\PastRB}{\textsl{Past}(\RB)}

\section{Introduction}
In light of the Wightman reconstruction theorem\cite{SW}, a presentation of a
relativistic quantum field theory just has to fix all the expectation values of
the vacuum state.
The most direct presentation in such terms would be to fix the expectation values of
the vacuum state directly, but most presentations are quite remote from the expectation
values.
We will discuss in section \ref{QKG}, therefore, an intermediate and very succinct
algebraic presentation of the quantized Klein-Gordon field.

Then in section \ref{CKG}, we construct a closely similar noncommutative algebraic
presentation of a quantum field that has the same probability density over its
configuration space as a classical Klein-Gordon random field at non-zero temperature.
In these algebraic presentations, a quantum field theory is very much closer to a
classical random field theory than quantum mechanics is to a classical particle
mechanics, which allows an understanding of quantum theory in terms of classical
\emph{random} fields.

In the view taken here, a classical description can be understood to be what we would
observe if our measurement apparatus were not affected at all by quantum fluctuations,
whereas a quantum description can be understood to be what we would observe if our
measurement apparatus were affected by quantum fluctuations to the same extent as
the rest of the universe (as in fact it is).
Finally, therefore, a heterodox quantum description is introduced in section
\ref{Intermediate} in which our measurement apparatus is less affected by quantum
fluctuations than the rest of the universe.
The principal technical characteristic of this interaction-free quantum field
theory is that its Lorentz invariant quantum vacuum state is not annihilated
by the annihilation operators of the theory, but is a Lorentz invariant analogue
of a thermal state.

\section{A succinct presentation of the quantized Klein-Gordon field}
\label{QKG}
Suppose that $\hat\phi:f\mapsto\hat\phi[f]$ is a linear operator valued map from a suitable
space of functions.
Typically $f$ is taken from a Schwartz space of functions\cite{Haag}, so that
$f(x)$ is infinitely often differentiable and decreases as well as its derivatives faster
than any power as $x$ moves to infinity in any direction.
$\hat\phi(x)$ is referred to as the operator valued distribution that generates $\hat\phi[f]$,
\begin{equation}
  \hat\phi[f]=\int \Intd^4x f(x)\hat\phi(x).
\end{equation}
Project $\hat\phi[f]$ into two parts,
\begin{equation}
  \hat\phi[f]=\hat a^\dagger[f]+\hat a[f],
\end{equation}
and specify the algebraic properties of $\hat a^\dagger$ and $\hat a$ by the commutation relations
\begin{equation}
  \Bigl[\hat a[g],\hat a^\dagger[f]\Bigr]=(f,g),\qquad
  \Bigl[\hat a[f],\hat a[g]\Bigr]=0,\qquad
  \Bigl[\hat a^\dagger[f],\hat a^\dagger[g]\Bigr]=0.
\end{equation}
The manifestly invariant Hermitian inner product $(f,g)$ is given by
\begin{equation}\label{InnerProduct}
  (f,g)=\hbar \int\frac{\Intd^4k}{(2\pi)^4}2\pi
         \delta(k^\mu k_\mu-m^2)\theta(k_0)\tilde f^*(k)\tilde g(k).
\end{equation}
This fixes the algebraic structure of the operators $\hat\phi[f]$.
All that remains to fix the vacuum expectation values is to give the trivial action
of the operators $\hat a[f]$ on the vacuum state,
\begin{equation}
  \hat a[f]\left|0\right> = 0, \qquad \left<0\right|\hat a^\dagger[f]=0,
\end{equation}
and specify the normalization of the vacuum vector $\left<0\right.\!\left|0\right> = 1.$

That's it. To calculate any vacuum expectation value, apply the commutation relations
above repeatedly, eliminating any terms in which $\hat a[f]\left|0\right>$ or
$\left<0\right|\hat a^\dagger[f]$ appear, until we obtain a number.
This Lorentz invariant presentation can be thought of as a relatively direct generating
scheme for experimentally observable correlation functions.
It is equivalent to Lagrangian and other presentations of the quantized Klein-Gordon field.
After the event, the algebraic structure allows an interpretation of $\hat a^\dagger[f]$ and
$\hat a[f]$ as creation and annihilation operators.

\section{The classical Klein-Gordon field at non-zero temperature}
\label{CKG}
There is no need to \emph{justify} this quantum field as the quantization of the classical
Klein-Gordon field; indeed, to do so is counterproductive, because the properties of the
quantized and classical Klein-Gordon fields are quite different.
It is more helpful to compare the properties of the quantized Klein-Gordon field with
the properties of the classical Klein-Gordon random field at non-zero temperature.
The emphasis on classical \emph{random} fields at non-zero temperature is necessary to make
a relatively direct comparison with the quantized Klein-Gordon field, because a differentiable
classical field cannot describe quantum and thermal fluctuations.
A classical random field has a structure very similar to a quantum field:
following the approach above, a classical random field is a random variable valued map from a
suitable space of functions, $X:f\mapsto X[f]$, and we can introduce a random variable
valued distribution $X(x)$ that generates $X[f]$,
\begin{equation}
  X[f]=\int \Intd^4x f(x)X(x).
\end{equation}
Just as a quantum field state generates expectation values for quantum observables, so a
classical random field state generates expectation values for classical random variables.

For the equilibrium state of the classical Klein-Gordon random field at temperature $T$,
the probability of observing a configuration $\phi_t(x)$ at time $t$ is given by
\begin{eqnarray}\label{ClassicalEquilibriumProb}
\rho_E[\phi_t]&\eqN&\exp{[-H_C[\phi_t]/\kT]}\cr
\noalign{\vspace{3pt}}
        &=&\exp{\left[-\frac{1}{\kT} \int\frac{\Intd^3k}{(2\pi)^3}
            \frac{1}{2}\tilde\phi_t^*(k)(|k|^2+m^2)\;\tilde \phi_t(k)\right]},
\end{eqnarray}
where $\eqN$ indicates equality up to a normalization constant.
In contrast, for the vacuum of the quantized Klein-Gordon field, the probability of
observing a configuration $\phi_t(x)$ at time $t$ is given\cite{MorganWigner} by
\begin{equation}\label{VacuumProb}
\rho_0[\phi_t] \eqN \exp{\left[-\frac{1}{\hbar}
        \int\frac{\Intd^3k}{(2\pi)^3}
            \tilde\phi_t^*(k)\sqrt{|k|^2+m^2}\;\tilde \phi_t(k)\right]}.
\end{equation}
There is notably little difference between these probability densities, but it is
of course significant: $\kT$ is replaced by $\hbar$; the Galilean symmetry group is
replaced by the Poincar\'e symmetry group; $\Half(|k|^2+m^2)$ is replaced by
$\sqrt{|k|^2+m^2}$.
These changes can all be thought of as aspects of the group theoretical difference.

From the point of view of these probability densities alone, disregarding for the
moment that measurement theory is different for classical random fields and for
quantum fields, $\kT$ and $\hbar$ both determine the amplitude of fluctuations.
The different functional forms for thermal fluctuations and quantum fluctuations are
combined in a thermal state of the quantized Klein-Gordon field in the probability
density
\begin{equation}
  \rho_T[\phi_t]\eqN\exp{\left[-\frac{1}{\hbar}
        \int\frac{\Intd^3k}{(2\pi)^3}
            \tanh{\left(\frac{\hbar \sqrt{|k|^2+m^2}}{2\kT}\right)}
            \tilde\phi_t^*(k)\sqrt{|k|^2+m^2}\;\tilde \phi_t(k)\right]}
\end{equation}
of observing a configuration $\phi_t(x)$ at time $t$, in which the integrand interpolates
between the $\rho_E$ integrand at low wave numbers (if $m\ll\kT/\hbar$) and the $\rho_0$
integrand at high wave numbers (an analogue of this probability density is derived in
\cite{MorganWigner} as a ``trajectory Wigner function'', but this paper deliberately uses
the more accessible idea of probability densities because they are adequate for the
argument made here).
Thermal fluctuations and quantum fluctuations \emph{are} different, as they have to be
if we are to think about quantum field theory in terms of fluctuations.

The probability density aspect of the classical Klein-Gordon random field at non-zero
temperature can be presented in a quantum field theoretical way,
just by replacing equation (\ref{InnerProduct}) by
\begin{equation}\label{CKGInnerProduct}
  (f,g)_C=\kT \int\frac{\Intd^4k}{(2\pi)^4}
         \frac{2\pi\delta(k^\mu k_\mu-m^2)\theta(k_0)}{\frac{1}{2}k_0}\tilde f^*(k)\tilde g(k),
\end{equation}
so that the equilibrium state is the $\left|0\right>$ state of the resulting quantum theory.
Nonequilibrium states can be generated from the equilibrium state in the usual
quantum field theoretic way, by the action of $a^\dagger[f]$ on the $\left|0\right>$ state.
Equation (\ref{CKGInnerProduct}) does lead to the probability density $\rho_E[\phi_t]$, but
the term $\theta(k_0)$ explicitly restricts wave-numbers to positive frequency, which is
classically somewhat heterodox.
The explicit arrow-of-time term could be left out, but \emph{that} would be quantum
theoretically somewhat heterodox, since it corresponds to there being no lower bound
for the energy.

Equally, the probability density aspect of the quantized Klein-Gordon field can be
presented in a classical random field theoretical way, just by introducing a Hamiltonian
\begin{equation}\label{quantumH}
  H_Q[\phi_t]= \int\frac{\Intd^3k}{(2\pi)^3} \tilde\phi_t^*(k)\sqrt{|k|^2+m^2}\;\tilde \phi_t(k).
\end{equation}
At the simple level of probability densities over configuration space, where they can be
compared directly, we have been able to characterize the difference between the classical
Klein-Gordon random field and the quantized Klein-Gordon field very clearly, but this
interchange of presentations of the probability density aspect ignores a significant
difference between the concepts of measurement in classical and quantum theory.
As a presentation of the classical Klein-Gordon random field, equation (\ref{CKGInnerProduct})
introduces incompatibility of measurements, while, as a presentation of the quantized
Klein-Gordon field, a probability density $\exp{[-H_Q[\phi_t]/\hbar]}$ based on
equation (\ref{quantumH}) implicitly asserts that all measurements are compatible.

A commutative classical algebra of observables is of course not isomorphic to a
noncommutative quantum algebra of observables, so states over the two algebras cannot
in general be equivalent, but, just because the Wigner function for the vacuum state
of the quantized Klein-Gordon field is positive definite, we \emph{can} fix a classical
state for a classical Klein-Gordon random field to be that Wigner function.

\section{An intermediate measurement algebra}
\label{Intermediate}
In physical terms, equation (\ref{CKGInnerProduct}), as part of a quantum theoretical
system, makes all measurement devices subject to thermal fluctuations at a universal
temperature $T$, as well as the measured system, while equation (\ref{quantumH}),
as part of a classical theoretical system, makes measurement devices not subject to the
empirical universality of quantum fluctuations.
Measurement devices \emph{are} subject to quantum fluctuations, unless in the future we
find a way to reduce them, but we can nonetheless \emph{imagine} what we would observe
if we could eliminate quantum fluctuations.
Even if we decide that we cannot imagine so much, nonetheless we can model the quantum
fluctuations of our real measurement devices \emph{explicitly}, just as we usually model
the thermal fluctuations of our real measurement devices explicitly (if we have to, but
we ignore thermal fluctuations whenever they make no observable difference).

This analysis suggests that we can introduce a variant of quantum field theory in which
measurement devices are ``quantum-cooled'', but quantum fluctuations of the measurement
devices are not as entirely eliminated as they are in classical measurement devices.
The Hermitian inner product $(f,g)$ of equation (\ref{InnerProduct}) essentially encodes
the amplitude of the quantum fluctuations of both the measurement devices and of the
rest of the universe.
To separately describe the fluctuations of the measurement devices and the fluctuations
of the rest of the universe, we can construct a quantum field theory in which the
commutation relations are taken essentially to describe interactions between measurement
devices, and the quantum fluctuations of the rest of the universe have an independent scale.
Suppose, therefore, that equation (\ref{InnerProduct}) is replaced by 
\begin{equation}\label{InnerProductXi}
  (f,g)_\xi=\xi\hbar \int\frac{\Intd^4k}{(2\pi)^4}2\pi
         \delta(k^\mu k_\mu-m^2)\theta(k_0)\tilde f^*(k)\tilde g(k).
\end{equation}
where $\xi>0$ is a real number, which will be less than $1$ if we have successfully
quantum-cooled our measurement devices.
For the state annihilated by $a[f]$, the probability of observing a configuration
$\phi_t(x)$ becomes
\begin{equation}\label{VacuumProbXi}
\rho_0^\xi[\phi_t] \eqN \exp{\left[-\frac{1}{\xi\hbar}
        \int\frac{\Intd^3k}{(2\pi)^3}
            \tilde\phi_t^*(k)\sqrt{|k|^2+m^2}\;\tilde \phi_t(k)\right]},
\end{equation}
so this is not the conventional vacuum state, with probability density $\rho_0[\phi_t]$,
but a different one, which is, nonetheless, Lorentz invariant.
To construct the conventional vacuum state, we modify the procedure used to construct a thermal
state, which invokes a Hamiltonian operator
\begin{equation}
  \hat H=\int\hat a^\dagger(k) \hat a(k) (k_0)^2 \frac{\Intd^3k}{(2\pi)^3},
\end{equation}
invoking instead, but in the same way, an operator
\begin{equation}
  \hat \Xi=\int\hat a^\dagger(k) \hat a(k) k_0 \frac{\Intd^3k}{(2\pi)^3},
\end{equation}
so that the expectation value for an observable $\hat A$ is given by
$\mathrm{Tr}\left[\exp{(-\hat\Xi/\lambda\hbar)\hat A}\right]$.
Then, using \cite[Appendix D]{MorganWigner}, the probability of observing a
configuration $\phi_t(x)$ is
\begin{equation}\label{VacuumProbXiLambda}
\rho_\lambda^\xi[\phi_t] \eqN \exp{\left[-\frac{1}{\xi\hbar}
        \int\frac{\Intd^3k}{(2\pi)^3}\tanh{\left(\frac{\xi}{2\lambda}\right)}
            \tilde\phi_t^*(k)\sqrt{|k|^2+m^2}\;\tilde \phi_t(k)\right]},
\end{equation}
so that, provided
\begin{equation}
   \lambda=\frac{\xi}{2\tanh{{}^{-1}(\xi)}},
\end{equation}
we again obtain $\rho_0[\phi_t]$.
This equation has solutions for $0<\xi<1$.
This conventional vacuum state, with an unconventional measurement algebra, is
essentially intermediate between a conventional quantized Klein-Gordon field and
a classical Klein-Gordon random field, in that the measurement algebra becomes
closer to the classical algebra of observables as $\xi\rightarrow 0$ (speaking
loosely, since the limit does not exist), while the observed state
is unchanged, in the sense that $\rho_\lambda^\xi[\phi_t]$ is unchanged
(at every time, and for every foliation).

Although this construction is instructive, it is more helpful to take either
conventional quantum field theory or classical random field theory as a
conceptual starting point for models of experimental apparatus.
For the above model to be useful, the detailed quantum fluctuations of different
measurement apparatuses would have to be describable by the same value of $\xi<1$,
which is unlikely to be the case.
Our measurement devices either will continue to be subject to universal quantum
fluctuations, so that a quantum field model will be appropriate, or else the detailed
quantum fluctuations of each measurement apparatus, subject to different levels
of quantum fluctuations, may be more easily described explicitly in a classical random
field model.

\section{Discussion}
\label{Discussion}
Remarkably from some points of view, the classical presentation of the probability density
aspect of the quantized Klein-Gordon field by equation (\ref{quantumH}) is manifestly
nonlocal\cite{SG}, explicitly exhibiting the nonlocality omnipresent in quantum mechanics
that has been identified by Hegerfeldt\cite{Hegerfeldt}, even though signal locality is
preserved.
It is noteworthy that Hegerfeldt nonlocality is quite different from the nonlocality usually
inferred from the experimental violation of Bell inequalities.
Bell inequalities cannot be derived for classical random fields unless assumptions are introduced
that are generally not satisfied if there are either thermal or quantum fluctuations\cite{MorganBell}.
From a classical analytical perspective, Hegerfeldt nonlocality can be understood as dynamical,
whereas the violation of Bell inequalities can be understood to be a result of an experimenter
making a special choice of initial conditions.

Bohr and Heisenberg abandoned a disturbance interpretation as an immediate and direct
result of the EPR paper in 1935, preferring an essentially positivistic
interpretation\cite{Fine}.
To my knowledge disturbance interpretations have not been directly suggested since
because of an apparently unreasonable nonlocality, although the idea of measurement
disturbance has never entirely gone away and can be found in modified form, for
example, in the \emph{unsharp} properties approach to Positive Operator Valued
Measures\cite{BGL}.
The approach of this paper, however, taken with Refs. \cite{MorganWigner} and
\cite{MorganBell}, allows a return to a disturbance interpretation of quantum theory,
provided it is in terms of classical random field models for complete experimental
apparatuses, not in terms of measurement of classical particle properties.
The possibility of interchanging classical and quantum presentations of quantum fields and
classical random fields, and the construction of quantum field theories that are in
a reasonable sense intermediate between them, offers new ways of thinking about both.

I am grateful for conversations with Piero Mana in Stockholm, whose idea it was to
write down equation (\ref{CKGInnerProduct}), and to a referee for helpful comments.

\end{document}